# Resilient Mobile Energy Storage Resources Based Distribution Network Restoration in Interdependent Power-Transportation-Information Networks

Jian Zhong, Chen Chen, *Senior Member, IEEE,* Qiming Yang, Dafu Liu, Wentao Shen, Chenlin Ji, and Zhaohong Bie, *Fellow, IEEE*

*Abstract*—**The interactions between power, transportation, and information networks (PTIN), are becoming more profound with the advent of smart city technologies. Existing mobile energy storage resource (MESR)-based power distribution network (PDN) restoration schemes often neglect the interdependencies among PTIN, thus, efficient PDN restoration cannot be achieved. This paper outlines the interacting factors of power supply demand, traffic operation efficiency, communication coverage, electric vehicle (EV) deployment capability, and PDN controllability among PTIN and further develops a PTIN-interacting model to reflect the 'chained recovery effect' of the MESR-based restoration process. On this basis, a two-stage PDN restoration scheme is proposed that utilizes three emergency resources, including EVs, mobile energy storage systems (MESSs), and unmanned aerial vehicles (UAVs), to restore the power supply and communication of PDNs. This scheme first improves the distribution automation function, EV deployment capability, and traffic operation efficiency by prioritizing the recovery of communication network (CN) and urban traffic network (UTN) loads. Then, EVs and MESSs are further scheduled to achieve a better PDN restoration effect with the support of the restored CNs and UTNs. Case studies on a PDN, CN, and UTN integrated test system are conducted to verify the effectiveness of the proposed scheme. The results show that the prioritized load recovery operation for CN and UTN facilities in this scheme greatly improves the PDN restoration effect.**

*Index Terms*—**Power, transportation, and information network interaction, vehicle-to-grid (V2G), distribution network restoration, cyber-physical interdependence.**

## NOMENCLATURE

*A. Indices and Sets*

| | |
|---|---|
| $a,b$ | Index for CN nodes |
| $i, j$ | Index for PDN buses |
| $k$ | Index for PDN lines |
| $l$ | Index for TLs in UTN |
| $m,n$ | Index for TJs in UTN |
| $r, s$ | Index for UAVs |
| $z$ | Index for EVs and MESSs |
| $\mathcal{F}_i^P$ | Set of PDN lines connected to node $i$ |
| $\mathcal{F}_{i^g}^U$ | Set of UAV sites in the recovery step $i^g$ |
| $\mathcal{L}^C$ | Set of CN links |
| $\mathcal{L}^P$ | Set of PDN lines |
| $\mathcal{L}^T$ | Set of UTN lanes |
| $\mathcal{N}^C$ | Set of CN nodes |
| $\mathcal{N}^P$ | Set of PDN nodes |
| $\mathcal{N}^T$ | Set of UTN nodes |

*B. Parameters*

| | |
|---|---|
| $x_a^C$,$y_a^C$ | Coordinates of CN node $a$ |
| $x_i^P$ ,$y_i^P$ | Coordinates of PDN bus $i$ |
| $x_m^T$,$y_m^T$ | Coordinates of TJ $m$ |
| $x_z^{EV}$ ,$y_z^{EV}$ | Coordinates of EV $z$ |

*C. Variables*

| | |
|---|---|
| $c_a^C$ | Communication state of CN node $a$ |
| $c_i^P$ | Communication state of PDN bus $i$ |
| $c_z^E$ | Communication state of EV $z$ |
| $e_a^C$ | Energized state of CN node $a$ |
| $e_i^P$ | Energized state of PDN bus $i$ |
| $e_m^T$ | Energized state of TJ $m$ |
| $e_l^{TL}$ | Energized state of TL $l$ |
| $g_{a,b}^C$ | Whether CN link $a$-$b$ is used |
| $g_i^{CP,a}$ | Whether the CN link between PDN bus $i$ and CN node $a$ can be used |
| $g_z^{CE,a}$ | Whether the CN link between EV $z$ and CN node $a$ can be used |
| $h_{a,b}^C$ | Whether CN nodes $a$ and $b$ are within the communication coverage |
| $h_i^{CP,a}$ | Whether PDN bus $i$ is covered by CN node $a$ |
| $h_z^{CE,a}$ | Whether EV $z$ are covered by CN node $a$ |
| $o_a^{O,r}$, $o_a^{D,r}$ | The battery power of UAV $r$ leaving from and arriving at site $a$ |
| $t_a^{O,r}$, $t_a^{D,r}$ | Time of UAV $r$ leaving from and arriving at site $a$ |
| $u_{a,b}^r$ | 0-1 variable indicating whether the path of $a-b$ is selected by UAV $r$ |
| $x_r^U$, $y_r^U$ | Coordinates of UAV $r$ |
| $w_a^{O,r}$, $w_a^{D,r}$ | 0-1 variable indicating whether UAV $r$ leaves from and arrives at site $a$ |
| $w_a^{R,z}$ | 0-1 variable indicating whether vehicle $z$ is dispatched to the V2G station $a$ |

## I. INTRODUCTION

THE frequent occurrence of natural disasters has caused widespread power outages in recent years, resulting in thousands of billions of dollars in economic losses [1]. Most of the outages were caused by faults in the power distribution networks (PDNs) [2]. Therefore, it is crucial to propose effective PDN service restoration schemes that ensure the continuity of the power supply [3] and minimize the impact of disasters on critical infrastructure [4]. However, extreme events may cause the PDN to be disconnected from the bulk power systems, and in this case, restoration will rely on distributed generations and energy storages to serve loads [5]. With the proliferation of electric mobility [6], mobile energy storage resources (MESRs), including electric vehicles (EVs) and mobile energy storage systems (MESSs) are rapidly developing in urban areas [7] and are proposed as a valuable backup in the event of major power outages [8].

Various types of MESRs have been explored for PDN resilience improvement. The potential of EVs for service restoration has been investigated by Sharma *et al.* [9]. Li *et al.* proposed a vehicle-to-grid (V2G) scheduling approach that takes into account the predicted V2G capacity [10]. To achieve improved resilience at the feeder level in an active PDN, Jamborsalamati *et al*. proposed an autonomous load restoration architecture based on EV aggregations [11]. Furthermore, electric buses are considered a type of high-performance EV resource for critical loads during extreme events in [4], [12], and [13]. In [14] and [15], Lei and Kim demonstrated the advantage of MESSs in improving the resilience of PDNs. However, current studies have not well considered the abovementioned mobile resources' dependence on information and transportation networks to participate in PDN restoration. This may make them inefficient in the load recovery process.

With the wide application of smart city technologies, the current power, transportation, and information networks (PTIN) have been upgraded from independent networks to more intelligent but interdependent networks [16]-[17]. Due to this integration property, the failure of a node in one network can lead to the failure of dependent nodes in other networks [18], resulting in a recursive cascading of failures across multiple interdependent networks [19]. Therefore, as shown in Fig. 1, the communication network (CN) and the urban transportation network (UTN) are deeply involved in the process of restoring power to loads using MESRs after a disaster in PDNs. Power outages can cause the loss of power supply to transportation facilities and communication facilities. On the one hand, the distribution automation (DA) of PDNs and the dispatching of EVs rely on normal working CNs. On the other hand, the loss of transportation facilities may lead to severe traffic congestion [20]. The MESR-based restoration process of the PDN involves the timely scheduling of mobile emergency resources in the UTN [21], which can easily be hindered by congestion. These interaction impacts could prolong the restoration process [22] and are not well considered in existing works. Although the interdependency of UTNs, CNs, and PDNs has been considered by many researchers in recent years, there is no complete modeling work for MESR-based restoration in interdependent PTIN due to the complexity of the interactions among the three networks and the difficulty of modeling the interactive recovery process. As PTIN coupling deepens, it is essential to build the interdependent PTIN model and propose corresponding resilience improvement schemes.

We believe that modeling the interdependent PTIN, on which PDN restoration is based, is a research gap. In addition, the integrated MESR-based PDN restoration scheme with multiple resource scheduling has not been well explored. Therefore, this paper proposes and quantifies the interacting factors of power supply demands, traffic operation efficiency, communication coverage, EV dispatching capability, and PDN controllability among PTIN. In this way, a PTIN interreacted model in which the dependences of MESR-based PDN restoration on UTN and CN facilities are specified is further developed. Based on this model, the scheduling scheme for three mobile emergency resources, including MESSs and EVs for power recovery, and unmanned aerial vehicles (UAVs) equipped with base stations for communication recovery, is proposed. This scheme reduces traffic congestion, improves EV scheduling capability, and enhances PDN DA function by prioritizing the restoration of power to CN and UTN facilities. This allows for better PDN restoration with the support of condition-improved CN and UTN. The main contributions of this paper are as follows:

1. The interacting factors of PTIN in MESR-based PDN restoration are identified and quantified. Then, the interdependent PTIN model is formed, reflecting the ‘chained recovery effect’ of the three networks during the PDN restoration process.

2. Mobile emergency resource scheduling models are built to help restore the PDN topology control and power supply capabilities. More realistic CN connections and UAV battery constraints are considered in these models, making them more practical in applications.

3. A two-stage post-disaster restoration scheme is proposed, which gives priority to the restoration of UTN and CN loads to reduce traffic congestion and improve the dispatching capability of EVs as well as the PDN control capability. Better PDN restoration is further achievable in the second stage of this scheme. Through a numerical case with PTIN-integrated networks, the potential of improving PDN resilience by prioritizing the recovery of UTN and CN loads is verified.

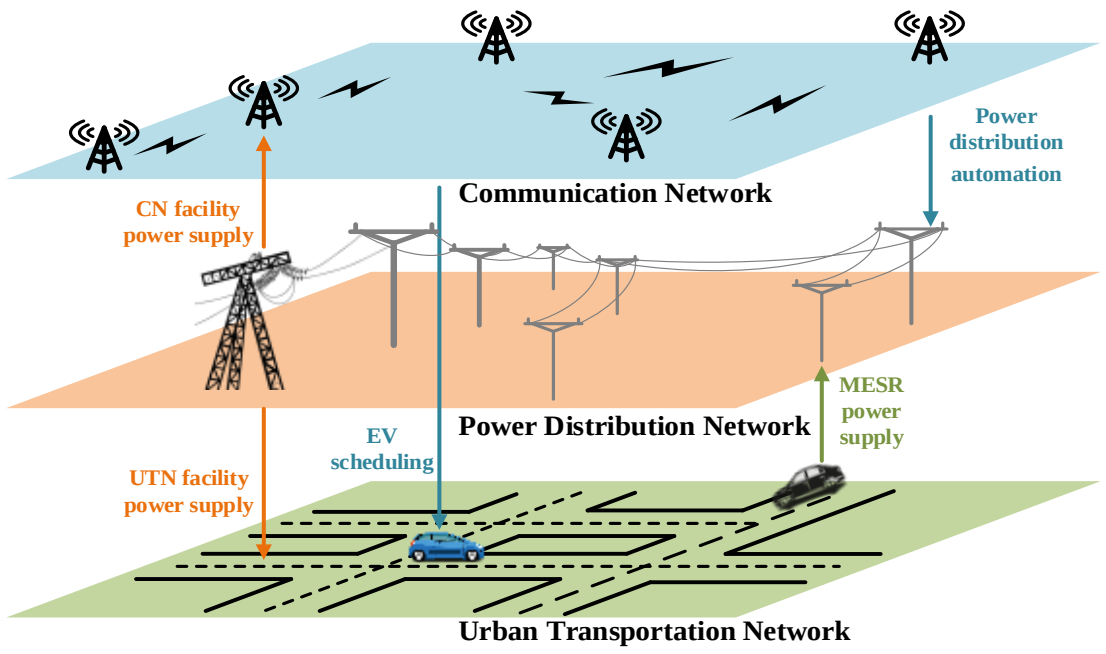

Fig. 1. The interdependency relationship among the PDN, UTN, and CN.

The rest of this paper is organized as follows. The framework and interactive factors of PTIN are introduced in Section II. Section III formulates the interdependent PTIN model and mobile emergency resource scheduling models, while Section IV provides the corresponding two-stage coordinated restoration schemes. The case studies and conclusions are presented and discussed in Sections V and VI, respectively.

## II. Coordinated Restoration Framework

The operational characteristics of PDNs, CNs, and UTNs must be modeled as a basis for exploring PDN restoration in interacted PTIN. Since wireless communication is widely used for DA [32] and EV scheduling [33], the CNs in this paper are also based on these wireless technologies. Meanwhile, the interdependence of the PTIN and mobile emergency resource scheduling constraints should be included in the restoration work.

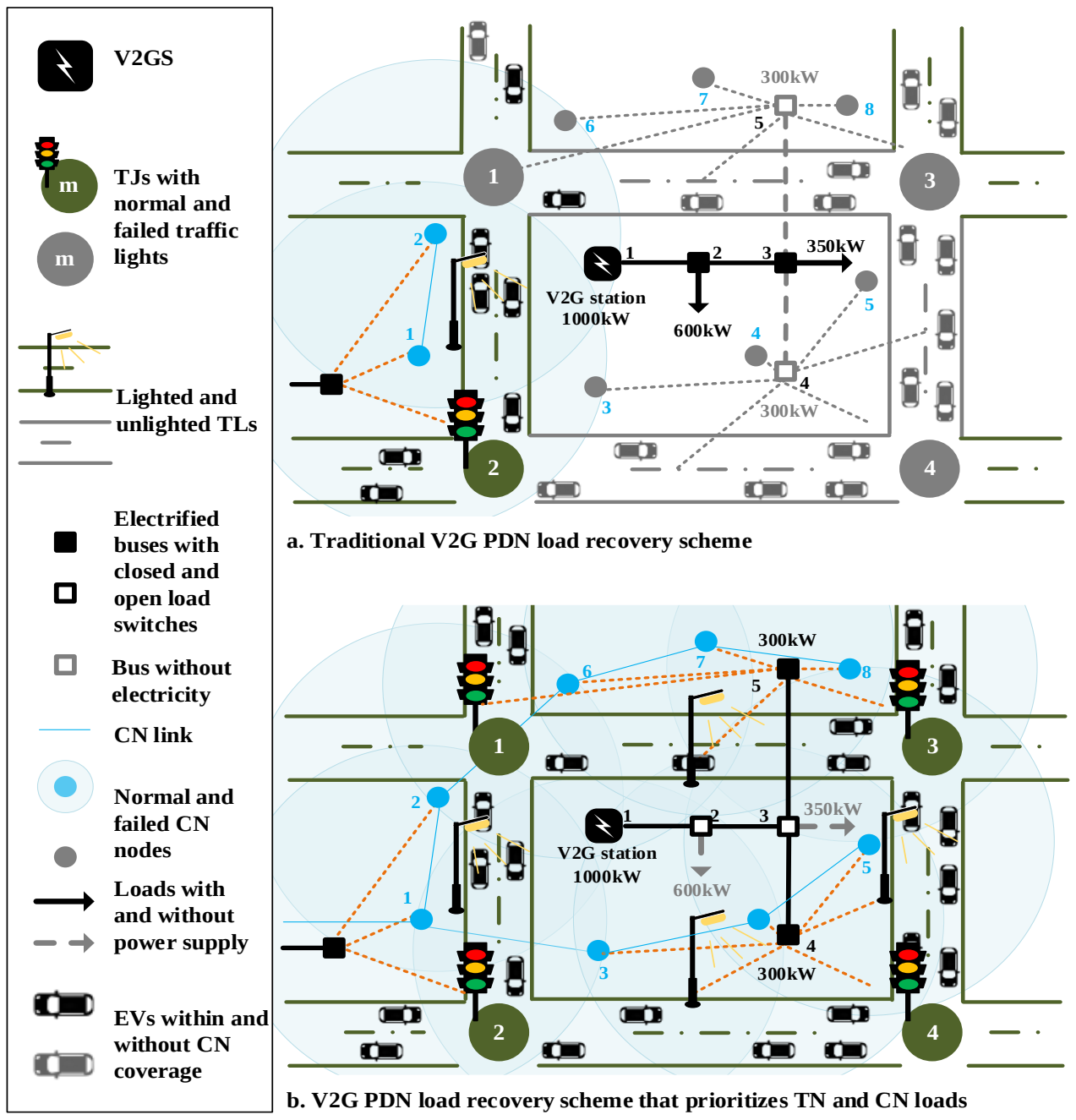

Fig. 2. A 'Chained reacted' PDN load recovery case.

A simple example with a V2G station (V2GS) is used to illustrate the interacting factors and 'chained recovery effect' among PDNs, CNs, and UTNs. As shown in Fig. 2, the V2GS has 1000 kW output power supported by EVs for load recovery. Power outages cause the loss of power supply to a lot of communication and transportation facilities. The DA of PDNs relies on CNs for the transmission of control commands from the operation center to feeder terminal units (FTUs) associated with buses to control the automated switches. When base stations 3-8 stop working due to the out of power supply, the automated switches belonging to PDN buses 2-5 will not be controllable. Moreover, most of EVs are out of communication coverage (the blue circles) and cannot be dispatched by the dispatching center, reducing the restoration power supply capacity. Power outages also cause the loss of power supply to transportation facilities including illumination lamps and traffic signal lights. First, the loss of illumination on traffic lanes (TLs) 2-4, 3-4, and 1-3 greatly affects the driving speed of roads in most low-visibility weather, which is often accompanied by extreme events. Second, traffic signal lights play an important role in traffic management in UTNs [16]. Due to self-interested driving behavior [19], the power loss of traffic signal lights at traffic junctions (TJs) may lead to severe UTN congestion [20]. The MESR-based restoration process of the PDN involves the timely scheduling of MESRs, which is easily hindered by congestion. This phenomenon prolongs the restoration process.

If power load recovery is performed in an upstream to downstream sequence in the traditional load recovery schemes as shown in Fig. 2(a), the loads on buses 2 and 3 are restored first. In this case, buses 4 and 5, on which most of the TN and CN loads depend, cannot be restored. In this way, the abovementioned phenomenon is not mitigated. However, as shown in Fig. 2(b), when the UTN and CN loads are restored with priority, both the traffic conditions and the communication coverage are improved. This recovery of CN achieves better DA function and more dispatchable EVs within the communication coverage. Meanwhile, the recovery of UTN establishes better traffic order. Then, more EVs can be dispatched through a UTN with better traffic conditions, allowing for faster and greater load recovery.

The above process involves power and communication recovery. Therefore, in this paper, three mobile emergency resources, EVs, MESSs, and UAVs, are utilized. Considering the willingness of personnel [30], only part of the EVs within the communication coverage will participate in the restoration process. MESSs are resources to restore the power supply to part of the crucial loads of the PDN after a disaster. It is assumed that MESSs are equipped with professional drivers and satellite phones parked in the warehouse and can be scheduled to V2GSs in complex traffic conditions. In the aftermath of a disaster, to pass through congested TLs and TJs, MESSs can request traffic police to guide them. UAVs equipped with base stations are demonstrated as promising emergency communication resources [24] that can spread information without being affected by the availability of roads and traffic congestion [23]. The UAVs are also set to be stocked in warehouses that provide battery replacement service, and can be deployed to designated locations for PDN facility communication recovery. In addition, the battery capacity constraint of UAVs is considered in the UAV scheduling scheme. In this paper, the communication recovery method is to deploy a limited number of UAVs to establish communication links from FTUs to the central nodes of CNs [28]. This is a more practical communication recovery scheme. When the communication is recovered, the automated switches, which are under the control of the corresponding FTUs, can be controlled by the operation center to open and close PDN lines.

The above case illustrates the benefits of prioritizing the power recovery to communication and transportation facilities during PDN restoration. Therefore, we propose a two-stage PDN restoration scheme. In the aftermath of a disaster, priority is given to restoring transportation and communication facility loads using dispatchable resources in the first stage. The objective of the first-stage algorithm is to restore the power supply to communication and transportation facilities using MESS, EV, and UAV resources. In this way, the operational efficiency of UTNs, EV dispatching capability, and DA function are improved. Then, a second-stage MESR dispatching algorithm is performed to restore the power supply to the PDN loads as much as possible with the support of the condition-improved CN and UTN. In this way, better PDN restoration can be achieved. The recovery process involves an interactive process among PDNs, CNs, and UTNs, along with vehicle routing in scenarios of changing traffic conditions. This makes the process hard to be accurately described by explicit constraints and needs to be represented by time-varying traffic flow simulation. Therefore, we decouple the two-stage recovery problem of the interdependent PTIN model into a multiple module cooperative process. Based on the existing communication coverage and deployable UAVs, two heuristic algorithms are formed to determine PDN restoration steps among recoverable PDN buses and lines in the above two restoration stages. After the above algorithms are solved, MESR and UAV scheduling models are called to derive the optimal resource allocation paths. Using the above method, the post-disaster recovery plan can be quickly generated. The above recovery scheme is summarized as the flowchart in Fig. 3.

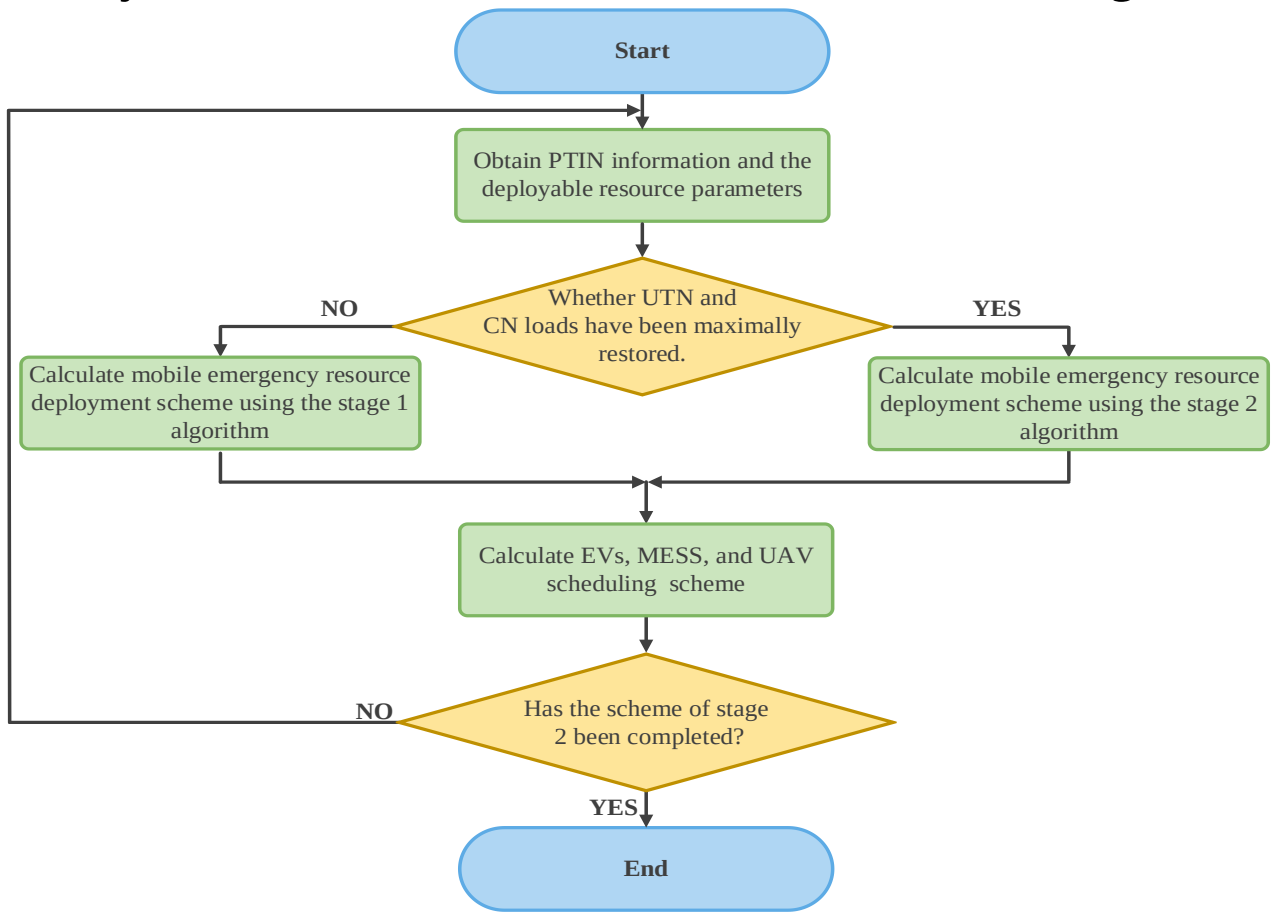

Fig. 3. Flowchart of the two-stage recovery scheme

## III. Problem Formulations

This section introduces the key models of the three networks and describes the interactive constraints among them. Then, the models for MESSs, EVs, and UAVs as emergency recovery resources for post-disaster restoration with their scheduling functions are presented.

### *A. Transportation Network Topology Modeling*

The UTN is represented by a connected directed graph $\mathcal{G}^T = [\mathcal{N}^T,\ \mathcal{L}^T]$. $\mathcal{N}^T$ is a set of sequentially numbered TJs, indexed by $m$ and $n$. $\mathcal{L}^T$ is a set of sequentially numbered TLs, indexed by $l$. A TJ is thought to be the origin or the terminus of any junction in a UTN, while a TL is an entity connecting two TJs. TLs are installed with lamps to illuminate streets, and TJs are installed with traffic signals to control vehicles to pass through in an orderly manner. Binary variables $e_m^T$ and $e_l^{TL}$ indicate the energized state of the facility at TJ $m$ and TL $l$, respectively, with '1' indicating energized and '0' indicating not energized.

*1) Traffic Network Constraints:* A TL $l$ is divided into $\chi^l$ interval sections indexed by $i^l$, which is described as

$$\lambda_l = \sum_{i^l=1,2,\ldots\ldots,\chi^l} \beta_{i^l}, \forall l \in \mathcal{L}^T, \quad (1)$$

$$v_{i^l}^{LS} = \beta_{i^l} / \tau_{i^l}^{LS}, \quad (2)$$

$$\tau_l^L = \sum_{i^l=1,2,\ldots\ldots,\chi^l} \tau_{i^l}^{LS}, \forall l \in \mathcal{L}^T, \quad (3)$$

where parameters $\tau_{i^l}^{LS}$ and $v_{i^l}^{LS}$ denote the average driving duration time and driving speed, respectively, of vehicles passing through the section $\beta_{i^l}$; $\tau_l^L$ and $\tau_m^J$ denote the whole driving duration time of the TL $l$ and TJ $m$, respectively, which are used to determine the fastest driving path in route planning.

*2) Vehicle Speed Constraints:* Each TL l and TJ m have their speed limits $v_l^{Lmax}$ and $v_m^{Jmax}$ depending on their traffic conditions. Vehicles should be driven at speeds below the corresponding speed limits. That is,

$$v_z^{E,i^l} \le v_l^{Lmax}, l \in \mathcal{L}^T, \forall z \in \mathcal{Z}^E, \quad (4)$$

$$v_z^{E,m} \le v_m^{Jmax}, m \in \mathcal{N}^T, \forall z \in \mathcal{Z}^E, \quad (5)$$

where $v_z^{E,i^l}$ and $v_z^{E,m}$ denote the driving speed of EV $z$ through the $i^l th$ section of TL $l$ and TJ $m$, respectively.

It is assumed that there are $\mathbb{N}^M$ MESSs included in a set $\mathcal{Z}^M$. Considering the above situation, the driving speed limitations of the MESS on TJs and TLs can be expressed as

$$v_z^{M,i^l} = (1 + \varpi_l^{LP}) \cdot v_{i^l}^{LS}, \forall z \in \mathcal{Z}^M, \quad (6)$$

$$v_z^{M,m} = \left(1 + \varpi_m^{JP}\right) \cdot v_m^J, \forall z \in \mathcal{Z}^M, \quad (7)$$

where $\varpi_l^{LP}$ and $\varpi_m^{JP}$ denote the increased speed rate at which traffic police give a MESS the right-of-way on TL $l$ and TJ $m$, respectively. $v_m^J$ denotes the average driving speed of vehicles passing through the section $\beta_{i^l}$. Vehicles traveling along a path from the origin TJ $m$ to the destination TJ $n$ are termed origin-destination (O-D) pairs $m - n$. For a certain O-D pair, there could be multiple paths. We denote $\mathcal{D}_{m,n}^{min}$ as the minimum driving time path set with minimum travel time $\tau_{m,n}$ for O-D pair $m - n$. In real UTNs, this driving path and speed information can be obtained from the intelligent traffic system (ITS). The software used for UTN simulation in this research, Simulation of Urban MObility (SUMO), is a discrete-time microscopic traffic simulator that provides simulation results consistent with real-world scenarios [25]-[26].

### *B. Power Distribution Network Topology Modeling*

The PDN is represented by a connected directed graph $\mathcal{G}^P = [\mathcal{N}^P,\ \mathcal{L}^P]$. $\mathcal{N}^P$ is the set of sequentially numbered buses, indexed by $i$ and $j$. $\mathcal{L}^P$ is the set of sequentially numbered lines, indexed by $k = (i, j)$ to indicate that the line $k$ is between buses $i$ and $j$. The V2GSs installed at certain buses are denoted by the set $\mathcal{S}^P \subset \mathcal{N}^P$. Binary variables $e_i^P$ and $c_i^P$ denote the energize state and communication state of the facilities of bus $i$, respectively. Parameters $\vartheta_i^P$ and $\vartheta_k^{PL}$ denote the equipment condition state (normal or breakdown) of the facilities of bus $i$ and line k, respectively. Binary variable $d_k^{PL}$ denotes the open and close control decisions of line $k$. The operation constraints and the single-commodity-flow-method-based topology constraints of PDNs are used as defined in [27].

*DA Control Constraints (DCC):* The control operation of a PDN line requires the consideration of its initial state. To ensure the safety of PDNs, lines in the open state can only be closed if both end buses of this line can communicate normally with the operation center. When controlling the closing of a line $k$, it is only necessary to establish communication with the FTU at the bus to which the automated switches belong. For a line $k = (i, j)$, two binary parameters, $\varrho_i^k$ and $\varrho_j^k$, denote which bus controls the switching of line $k$. Then, the PDN line control constraint can be expressed as

$$\left[\zeta_k^{PL} - \left(\varrho_i^k c_i^P + \varrho_j^k c_j^P\right)\right] \le d_k^{PL} \le \left[\zeta_k^{PL} + 0.5 \cdot \left(c_i^P + c_j^P\right)\right], \quad (8)$$

where $\zeta_k^{PL}$ denotes the initial open and closed state of line $k$. The above constraints simultaneously satisfy the opening and closing function for multiple PDN lines with a single automated switch or switches at both ends. Each bus has a load switch between the bus and its load, that is,

$$d_i^{Load} \le e_i^P, \forall i \in \mathcal{N}^P, \quad (9)$$

where variable $d_i^{Load}$ denotes the load switch state (open or closed) for bus $i$; The load switch is automatically disconnected when its corresponding bus is not energized.

### *C. Communication Network Topology Modeling*

The CN is represented by a connected directed graph $\mathcal{G}^C = [\mathcal{N}^C, \mathcal{L}^C]$. $\mathcal{N}^C$ is the set of sequentially numbered $\mathbb{N}^C$ communication nodes (i.e., base stations and UAVs equipped with base stations), indexed by $a$ and $b$. $\mathcal{L}^C$ is the set of communication links among CN nodes. The central network nodes are interfaces to connect the entire CN to the Internet [28], [29], denoted by the set $\mathcal{W}^C \subset \mathcal{N}^C$. Then, we assume that a CN is established as a radial topology with links among nodes. Thus, the topology is also built using the single-commodity-flow method. Only nodes that are connected to the central network nodes through normal links and nodes are judged to have normal communication states.

*1) Communication Coverage Constraints (CCC):* Setting the effective transmission distance of CN node $a$ as $\rho^a$, the premise that the communication link among nodes $a$ and $b$ ($a \neq b$) can be described by the following constraints:

$$h_{a,b}^C = h_{a,b}^C = \varepsilon(\rho^{a,b} - \sqrt{(x_a^C - x_b^C)^2 + (y_a^C - y_b^C)^2}, \quad (10)$$

$$g_{a,b}^C \leq h_{a,b}^C, \forall a, b \in \mathcal{N}^C, \quad (11)$$

$$g_{a,b}^C = g_{b,a}^C, \forall a, b \in \mathcal{N}^C, \quad (12)$$

$$g_{a,a}^C = 0, \forall a \in \mathcal{N}^C, \quad (13)$$

where $\rho^{a,b}$ is the larger one of the ranges covered by the two CN nodes $a$ and b; binary variable $g_{a,b}^{CN}$ denotes whether the communication link between nodes $a$ and $b$ is used.

*2) Communication Topology Constraints (CTC):* The commodity flows on links should satisfy:

$$f_{a,b}^C + f_{b,a}^C = 0, \forall a, b \in \mathcal{N}^C, \quad (14)$$

where $f_{a,b}^C$ denotes the commodity flow among links $a - b$ from CN node $a$ to $b$. Note that a disconnected link has no commodity flow through it, which can be expressed as

$$-M \cdot g_{a,b}^C \leq f_{a,b}^C \leq M \cdot g_{a,b}^C, \forall a, b \in \mathcal{N}^C, \quad (15)$$

where a very large number $M$ is used to express the conditional statement. The communication states of both nodes are equal when a link is used, that is,

$$\left(1 - g_{a,b}^C\right) \cdot (-M) + c_a^C \leq c_b^C \leq \left(1 - g_{a,b}^C\right) \cdot M + c_a^C. \quad (16)$$

The sum of commodities flowing out of a node is zero, which is described by the following two equality constraints:

$$\textstyle\sum_{b \in \mathcal{N}^C} f_{a,b}^C + c_a^C = 0, \forall a \in \mathcal{N}^C \backslash \mathcal{W}^C, \quad (17)$$

$$\textstyle\sum_{b \in \mathcal{N}^C} f_{a,b}^C + c_a^C = f_a^{CO}, \forall a \in \mathcal{W}^C, \quad (18)$$

where variable $f_a^{CO}$ denotes the commodity flow injected by the central network node $a$.

Additionally, the radial topology constraint is introduced to prevent circulating flows, which is written as

$$\textstyle\sum_{a \in \mathcal{N}^C} c_a^C - \sum_{a \in \mathcal{W}^C} c_a^C = 0.5 \cdot \sum_{a \in \mathcal{N}^C} \sum_{b \in \mathcal{N}^C} g_{a,b}^C. \quad (19)$$

### *D. Interdependency Constraints of CNs, PDNs, and UTNs*

*1) Power Supply Impact Constraints:* TJ $m$ and TL $l$ have their corresponding power supply buses $i$ and $j$. CN node $a$ has its power supply bus $i$. Their normal functioning requires that the load switches of the corresponding buses are closed, that is,

$$e_m^T \leq d_i^{Load}, \forall m \in \mathcal{N}^T, \quad (20)$$

$$e_l^{TL} \leq d_j^{Load}, \forall l \in \mathcal{L}^T, \quad (21)$$

$$e_a^C \leq d_i^{Load}, \forall a \in \mathcal{N}^C. \quad (22)$$

*2) Communications Impact Constraints (CIC):* Considering the communication requirement of DA, FTUs are controllable when they are within the coverage of normally communicating CN nodes, which can be described by the following constraints:

$$\varepsilon\left(\rho^{a,i} - \sqrt{(x_i^P - x_a^C)^2 + (y_i^P - y_a^C)^2}\right) = h_i^{CP,a}, \forall a \in \mathcal{N}^C, \quad (23)$$

$$h_i^{CP,a} - 1 + c_a^C \leq g_i^{CP,a} \leq 1 - h_i^{CP,a} + c_a^C, \forall a \in \mathcal{N}^C, \quad (24)$$

$$\textstyle\sum_{a \in \mathcal{N}^C} g_i^{CP,a} / \mathbb{N}^C \leq c_i^P \leq \sum_{a \in \mathcal{N}^C} g_i^{CP,a}, \forall a \in \mathcal{N}^C. \quad (25)$$

EVs can only be dispatched to V2GSs if they are within the coverage of a normally communicating CN node, that is,

$$\varepsilon\left(\rho^{a,z} - \sqrt{(x_z^E - x_a^C)^2 + (y_z^E - y_a^C)^2}\right) = h_z^{CE,a}, \forall a \in \mathcal{N}^C, \quad (26)$$

$$h_z^{CE,a} - 1 + c_a^C \leq g_z^{CE,a} \leq 1 - h_z^{CE,a} + c_a^C, \forall a \in \mathcal{N}^C, \quad (27)$$

$$\textstyle\sum_{a \in \mathcal{N}^C} g_z^{CE,a} / \mathbb{N}^C \leq c_z^E \leq \sum_{a \in \mathcal{N}^C} g_z^{CE,a}, \forall a \in \mathcal{N}^C. \quad (28)$$

*3) Traffic Impact Constraints:* Loss of illumination on the lanes affects the maximum driving speed on TLs. Loss of traffic signal indication leads to traffic congestion in the area adjacent to TJs, which reflects a reduction in the driving speed. They are described by the following equality constraints:

$$v_l^{Lmax} = [e_m^T + (e_m^T - 1) \cdot \xi^l] \cdot v^{Lmax}, \forall l \in \mathcal{L}^T, \quad (29)$$

$$v_m^{Jmax} = [e_m^T + (e_m^T - 1) \cdot \xi^n] \cdot v^{Jmax}, \forall m \in \mathcal{N}^T, \quad (30)$$

where $v^{Lmax}$ and $v^{Jmax}$ are the prescribed speed limits of TLs and TJs. Since TLs are linked with TJs, congestion at TJs can further affect driving speeds on adjacent TLs.

### *E. MESR Scheduling Model*

*MESR Scheduling Objective Function (MSOF):* The output power $\sigma^z$ of dispatchable vehicle $z$ (an EV in communication coverage or a MESS) and its driving time to each V2GS can be acquired. Binary variables $w_{i^v}^{R,z}$ denote whether vehicle $z$ is dispatched to V2GS $i^v$. A vehicle can at most be dispatched to one V2GS. Coefficient $\eta$ indicates the probability of EVs participating in the V2G restoration [30]. Based on the derived power requirement $\varsigma^{i^v}$ for each V2GS, the objective function is formulated as:

$$\min t_{DR}^{min} \quad (31)$$

$$s.t. \textstyle\sum_{i^v = 1:\mathbb{N}^V} w_{i^v}^{R,z} \leq 1, \forall z \in (\mathcal{Z}^E \cup \mathcal{Z}^M), \quad (32)$$

$$\left(\eta \cdot \textstyle\sum_{z \in \mathcal{Z}^E} w_{i^v}^{R,z} \cdot \sigma^z + \sum_{z \in \mathcal{Z}^M} w_{i^v}^{R,z} \cdot \sigma^z\right) \geq \varsigma^{i^v}, \forall i^v \in \mathcal{S}^P, \quad (33)$$

$$w_{i^v}^{R,z} \cdot \tau_{i^v}^{R,z} \leq t_{DR}^{min}, \forall z \in (\mathcal{Z}^E \cup \mathcal{Z}^M), \quad (34)$$

where $t_{DR}^{min}$ denotes the minimum elapsed time for the recovery process and $\tau_{i^v}^{R,z}$ denotes the minimum time taken by vehicle $z$ to drive from its location to the V2GS $i^v$.

### *F. UAV Scheduling Model*

We consider that solving the deployment location problem of UAVs is to find a minimum time-consuming solution for $\mathbb{N}^U$ UAVs among $\epsilon$ ($\epsilon = 1 + \kappa + \phi$) deployment sites indexed by $a^S$ and $b^S$ included in set $\mathcal{S}^U$: a set-off site (warehouse site for setting off and returning), $\kappa$ UAV deployment sites, and $\phi$ battery replacement sites (warehouse sites for battery replacement trips). All warehouse sites are sorted into a Set $\mathcal{H}$.

*1) UAV Deployment Sites Solving Function (UDSSF):* Assume that there are $n$ UAVs sorted into a set $\mathcal{Z}^U$. We use a binary variable $d_r^U$ to indicate whether UAV $r$ is deployed to coordinates $(x_r^U, y_r^U)$ for PDN communication recovery. For a PDN line $k=(i,j)$, given the known initial state $\zeta_k^{PL}$ and the control requirements $b_k^{PL}$, the UAV deployment goal is to find the least number of UAV deployment sites to achieve control of PDN lines. The objective function is formulated as follows:

$$\min(\sum_{r\in\mathcal{Z}^U} d_r^U) \tag{35}$$

$$s.t. \text{CCC: (10)} - \text{(13)}, \text{CTC: (14)} - \text{(19)}$$

$$(1+\zeta_k^{PL}-d_k^{PL})\cdot(-M)+2 \le c_i^P + c_j^P \le (1+\zeta_k^{PL}-d_k^{PL})\cdot M+2, \tag{36}$$

$$(1-\zeta_k^{PL}+d_k^{PL})\cdot(-M)+\varrho_i^k+\varrho_j^k \le \varrho_i^k\cdot c_i^P+\varrho_j^k\cdot c_j^P \le (1-\zeta_k^{PL}+d_k^{PL})\cdot M+\varrho_i^k+\varrho_j^k. \tag{37}$$

Constraints (36) and (37) represent the DA communication requirement of open and close operations on PDN lines with the initial closure and open states, respectively.

*2) Site Selection Constraints (SSC):* Matric $U_r$ indicates the travel path selection of UAV $r$. $u^r_{a^s,b^s}$ is a binary variable that indicates whether the path from sites $a^S$ to $b^S$ is selected for UAV $r$. The corresponding equations are defined as

$$U_r=\begin{pmatrix} u^r_{1,1} & \cdots & u^r_{1,\epsilon} \\ \vdots & \ddots & \vdots \\ u^r_{\epsilon,1} & \cdots & u^r_{\epsilon,\epsilon}\end{pmatrix}, \forall r\in\mathcal{Z}^U, \tag{38}$$

$$u^r_{a^s,a^s}=0, \forall a^s\in\mathcal{S}^U, \tag{39}$$

$$u^r_{a^s,b^s}=0, \forall a^s\in\mathcal{H}, \forall b^s\in\mathcal{H}, \tag{40}$$

$$w^{O,r}_{a^s}=\sum_{b^s=1:(\kappa+\phi)} u^r_{a^s,b^s}, \forall a^s\in\mathcal{S}^U, \tag{41}$$

$$w^{D,r}_{a^s}=\sum_{b^s=1:(\kappa+\phi)} u^r_{b^s,a^s}, \forall a^s\in\mathcal{S}^U, \tag{42}$$

$$\sum_{r=1,\cdots,\mathbb{N}^U} w^{O,r}_{a^s}=\sum_{r=1,\cdots,\mathbb{N}^U} w^{D,r}_{a^s}=1, \forall a^s\le 1+\kappa, \tag{43}$$

$$w^{O,r}_{a^s}=w^{D,r}_{a^s}, \forall a^s\in\mathcal{S}^U. \tag{44}$$

The number of departures and arrivals of a UAV at a deployment site should be equal. There must be one arrival and one departure of the UAVs for each deployment site. UAVs should depart from the warehouse (the first site) at the beginning and return to the warehouse at the end of the process. The O-D of trips cannot be selected among battery replacement sites and the set-off site. Battery replacement trips can be selected according to the battery residue capacity of UAVs. However, the number of arrivals and departures of battery replacement sites should be equal. That is,

$$\sum_{r=1,\cdots,\mathbb{N}^U} w^{O,r}_{a^s}=\sum_{r=1,\cdots,\mathbb{N}^U} w^{D,r}_{a^s}\le 1, (1+\kappa<a^s\le\epsilon), \tag{45}$$

*3) Departure and Arrival Time Constraints (DATC):* Only the selected deployment sites have departure time and arrival time. The departure time and arrival time of UAV $r$ at the site $a^s$ should be:

$$0\le t^{O,r}_{a^s}\le w^{O,r}_{a^s}\cdot M, \forall r\in\mathcal{Z}^U, \tag{46}$$

$$0\le t^{D,r}_{a^s}\le w^{D,r}_{a^s}\cdot M, \forall r\in\mathcal{Z}^U, \tag{47}$$

where variables $t^{O,r}_{a^s}$ and $t^{D,r}_{a^s}$ represent the time of departure from and arrival at the site $a^s$, respectively, of UAV $r$.

The time for each UAV to reach the destination of a path $a^s-b^s$ (i.e., site $b^s$) equals the departure time from the origin (i.e., site $a$) plus the time it travels along the path. The constraint is expressed as:

$$-(1-u^i_{a^s,b^s})\cdot M+t^{D,i}_{b^s}\le t^{O,r}_{a^s}+\lambda^{a^s,b^s}/v^U\le(1-u^i_{a^s,b^s})\cdot M+t^{D,i}_{b^s}, \forall a^s,b^s\in\mathcal{S}^U, \tag{48}$$

where $\lambda^{a^s,b^s}$ denotes the distance between sites $a^s$ and $b^s$ and $v^U$ denotes the UAV flying speed.

Moreover, if UAV $r$ is deployed at the battery replacement site $a^s$, the sum of the arrival time and battery replacement time should be larger than the departure time at this site. That is,

$$-(1-w^{O,r}_{a^s})\cdot M+t^{D,r}_{a^s}\le t^{O,r}_{a^s}-\tau^U_{a^s}, \forall(1+\kappa)<a^s\le\epsilon, \tag{49}$$

where $\tau^U_{a^s}$ denotes the battery replacement time at the site $a^s$.

*4) Battery Capacity Constraints (BCC):* If UAV $r$ is deployed at site $a^s$, the battery residue capacity of UAV $r$ arriving at the site $a^s$ minus the consumed electricity working at the site $a^s$ should equal the residue capacity of UAV $r$ departing from the site $a^s$. For path $a^s-b^s$, the amount of battery power at the site $a^s$ minus the consumed electricity for flying along the path $a^s-b^s$ is equal to the amount of battery residue capacity reaching the site $b^s$. That is,

$$0\le o^{O,i}_{a^s}\le w^{O,r}_{a^s}\cdot M, \forall r\in\mathcal{Z}^U, \tag{50}$$

$$0\le o^{D,i}_{a^s}\le w^{D,r}_{a^s}\cdot M, \forall r\in\mathcal{Z}^U, \tag{51}$$

$$-(1-w^{O,r}_{a^s})\cdot M+o^{D,r}_{a^s}\le o^{O,r}_{a^s}+\rho^r_{a^s}\le(1-w^{O,r}_{a^s})\cdot M+o^{D,r}_{a^s}, \forall r\in\mathcal{Z}^U, \tag{52}$$

$$-(1-u^r_{a^s,b^s})\cdot M+o^{O,r}_{a^s}\le o^{D,r}_{b^s}+\lambda^{a^s,b^s}\cdot\delta^{UAV}\le(1-u^r_{a^s,b^s})\cdot M+o^{O,r}_{a^s}, \forall r\in\mathcal{Z}^U, \tag{53}$$

where variables $o^{O,r}_{a^s}$ and $o^{D,r}_{a^s}$ represent the battery residue capacity of UAV $r$ when departing from and arriving at the site $a^s$, respectively. The residue capacity of UAV batteries is 100% when leaving warehouse sites. Thus, we can express this constraint as follows:

$$-(1-w^{O,r}_{a^s})+1\le o^{O,r}_{a^s}\le(1-w^{O,r}_{a^s})+1, \ \forall(1+\kappa)<a^s\le\epsilon, \forall r\in\mathcal{Z}^U. \tag{54}$$

*5) Workgroup Coordination Constraints (WCC):* The UAV deployment sites in the $i^g th$ step are recorded as a workgroup $\mathcal{F}^U_{i^g}$. Each workgroup $i^g$ is a series of UAV deployment sites where several UAVs work simultaneously to establish communication among FTUs and base stations. A UAV should not work more than once in each workgroup. The departure time of the former group $i^g$ should be earlier than the arrival time of the latter group $j^g$. Moreover, the arrival time of each site in a group $i^g$ plus the DA working time $\tau^U_{i^g}$ should be earlier than the departure time of each site in this group. That is,

$$\sum_{r=1,\cdots,\mathbb{N}^U} t^{D,r}_{a^s}-\sum_{r=1,\cdots,\mathbb{N}^U} t^{O,r}_{b^s}\ge\tau^U_{i^g}, \forall a^s,b^s\in\mathcal{F}^U_{i^g}, \tag{55}$$

$$\sum_{r=1,\cdots,\mathbb{N}^U} t^{D,r}_{a^s}\ge\sum_{r=1,\cdots,\mathbb{N}^U} t^{O,r}_{b^s}, \forall a^s,b^s\in\mathcal{F}^U_{j^g}, \tag{56}$$

$$\sum_{r=1,\cdots,\mathbb{N}^U} w^{D,r}_{a^s}\le 1, \forall a^s\in\mathcal{F}^U_{i^g}. \tag{57}$$

*6) Topological Sequential Control Constraints (TSCC):* Considering the interreacted recovery in the collaborative restoration process, the temporal order of site recovery must be planned. The arrival time of a UAV should lag the time when the V2GS reaches the corresponding load demand, that is,

$$\sum_{r=1,\cdots,\mathbb{N}^U} t^{O,r}_{a^s}\ge t^{Sa}_{i^g}, \forall a^s\in\mathcal{F}^U_{i^g}, \tag{58}$$

where variable $t^{Sa}_{i^g}$ denotes the time when the V2GSs reach the load demand for the $i^g th$ step.

*7) UAV Scheduling Objective Function (USOF):* For cases where deployment points are known, UAV deployment aims to complete the whole DA operation work in the shortest time. Therefore, the objective function is defined as follows:

$$\min\ t^{min}_{DU} \tag{59}$$

$$s.t.\ \mathrm{SSC}\text{: (38)-(45), DATC: (46)-(49),}$$
$$\text{BCC: (50)-(54), WCC: (55)-(57), TSCC: (58)}$$
$$t_1^{D,r} - \sum_{a^s\in\mathcal{S}^U} u^r_{a^s,1}\cdot\lambda^{a^s,1}/v^U \le t_{DU}^{min}, \forall r\in\mathcal{Z}^M. \quad (60)$$

The UAV travel scheme can be solved using the above model. It is worth noting that in cases of large regional urban recovery with a lot of UAVs and communication failed PDN lines, the number of variables in the above matrix $U_r$ increases by a multiple. To solve this problem, the restoration area can be divided into multiple regions according to regional proximity and microgrid division structure, and the UAVs can also be divided into multiple groups to form the fastest scheduling scheme for each region separately.

## IV. Two-stage Restoration Algorithms

Assume that UAV patrol technology is used to obtain information about the states of PDNs, traffic conditions of UTNs, and communication states of CNs after a disaster. The two-stage recovery is based on the information acquired above.

### A. First-Stage Recovery Algorithm

The first step of the first-stage algorithm is to ensure that there are no connections among V2GSs, which facilitates the subsequent formation of a radial topology. Then, the algorithm evaluates the recoverable load of each V2GS based on the interaction effects of PDNs, UTNs, and CNs. A circle judgment is performed to determine the PDN lines that can be controlled to be electrified by V2GSs. Three sets $\Psi^l$, $\Theta^l$, and $\Xi^l$ are used to include the selected PDN lines, their loads, and their UAV deployment groups. During each round of the cycle, PDN buses that can be controlled are searched for and placed into the set of recoverable buses. The buses with the highest loads of communication and transportation facilities are selected after each round of the cycle, and only the load switches of the buses with UTN and CN loads are closed. The method is formed as the following algorithm.

**Algorithm 1** MESR and UAV deployment in the first stage

**Initialize** the PDN and UAV parameters, and the cyclic judgment parameter $\alpha = 1$;
**if** the PDN is not radial topology **then**
  Search for an appropriate line to open to form the PDN as a radial topology;
**end**
**while** $\alpha \neq 0$ **do**
  $\alpha = 0$ ;$\mathrm{K}^l = []$; $\Gamma^l = []$;$\Upsilon^l = []$; $\Lambda^l = []$;
  **for** $k = (i,j) \in \mathcal{L}^P$, **do**
    **if** $e_i^P + e_i^P = 1$ **then**
      **if** $c_i^P + c_j^P = 2$ **then**
        Add $k$ into the set $\mathrm{K}^l$, its load into the set $\Gamma^l$, its CN and UTN load into the set $\Upsilon^l$, $\alpha = \alpha + 1$;
      **end**
    **else**
      **if** $UDSSF(k)$ has a solution **then**
        Add $k$ into the set $\mathrm{K}^l$, its load into the set $\Gamma^l$, its CN and UTN load into the set $\Upsilon^l$, its UAV deployment site group into the set $\Lambda^l$, $\alpha = \alpha + 1$;
      **end**
    **end**
  **end**
  Select the line with the maximum load to be electrified from set $\Upsilon^l$, add this line, its load, and UAV sites into sets $\Psi^l$, $\Theta^l$, and $\Xi^l$;
**end**
**if** dispatchable power capacity is greater than load demand **then**
  Allocate MESRs to V2GSs according to load demand;
**else**
  Allocate MESRs according to load demand of the maximum recoverable steps;
**end**
**output** MESS, EV, and UAV deployment results.

Based on the sequential recovery results from this algorithm, the resource scheduling schemes are derived from the objective functions of $USOF$ and $MSOF$. The residual capacity of a V2GS is the total capacity of the MESRs in this station minus the loads it picked up. When the residual capacity of the V2GS reaches the load demand for the current recovery step, this step is performed.

### B. Second-Stage Recovery Algorithm

After restoring power to communication and transportation facilities in the first stage, TN and CN conditions gradually improve, and the number of EVs that can be deployed increases. Then, MESRs are allocated to V2GSs according to the following algorithm.

**Algorithm 2** MESR and UAV deployment in the second stage

**Initialize** the PDN and UAV parameters;
$\mathrm{K}^n = []$; $\Gamma^n = []$;$\Upsilon^n = []$; $\Lambda^n = []$;
**for** $i \in \mathcal{N}^P$, **do**
  **if** $e_i^P = 1$ and $b_i^{load} = 0$ **then**
    **if** $c_i^P = 1$ **then**
      Add $i$ into the set $\mathrm{K}^n$, its load into the set $\Gamma^n$, its CN and UTN load into the set $\Upsilon^n$;
    **else**
      **if** $UDSSF(k, \varrho_i^k = 1, \varrho_j^k = 0)$ has a solution **then**
        Add $i$ into the set $\mathrm{K}^n$, its load into the set $\Gamma^n$, its CN and UTN load into the set $\Upsilon^n$, its UAV deployment site group into the set $\Lambda^n$;
      **end**
    **end**
  **end**
**end**
Calculate the amount of recoverable load for each V2GS;
Determine the capacity of MESRs that can be deployed to V2GSs;
**if** dispatchable power capacity is greater than load demand **then**
  Allocate MESRs according to load demand;
**else**
  Allocate MESRs in proportion to load demand;
**end**
**output** MESS, EV, and UAV deployment results.

According to the results from the above algorithm, EVs, MESSs, and UAVs are also scheduled to recover power loads using $MSOF$ and $USOF$ objective functions. When the residual capacity of the V2GS reaches the maximum load it can pick up, and that load is recovered.

## V. Case Studies

### A. Simulation Parameters and Testbed

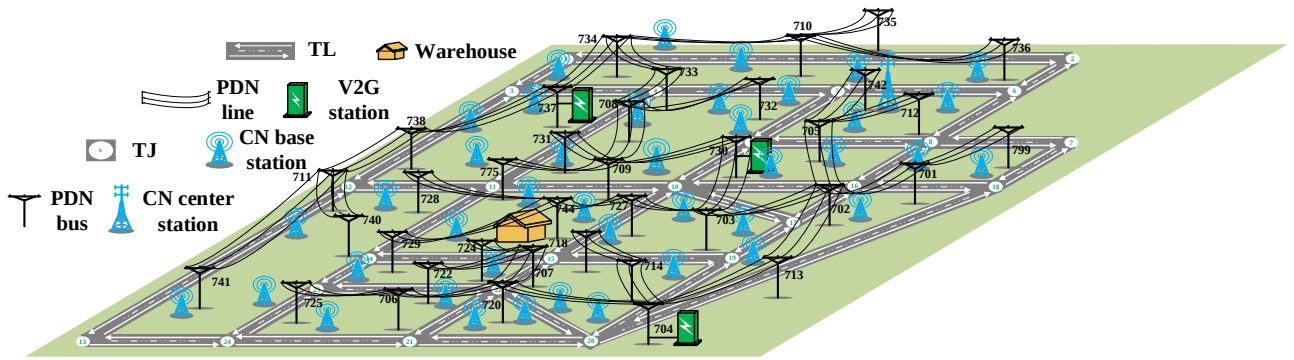

**Fig. 4.** A CN, UTN, and PDN integrated case

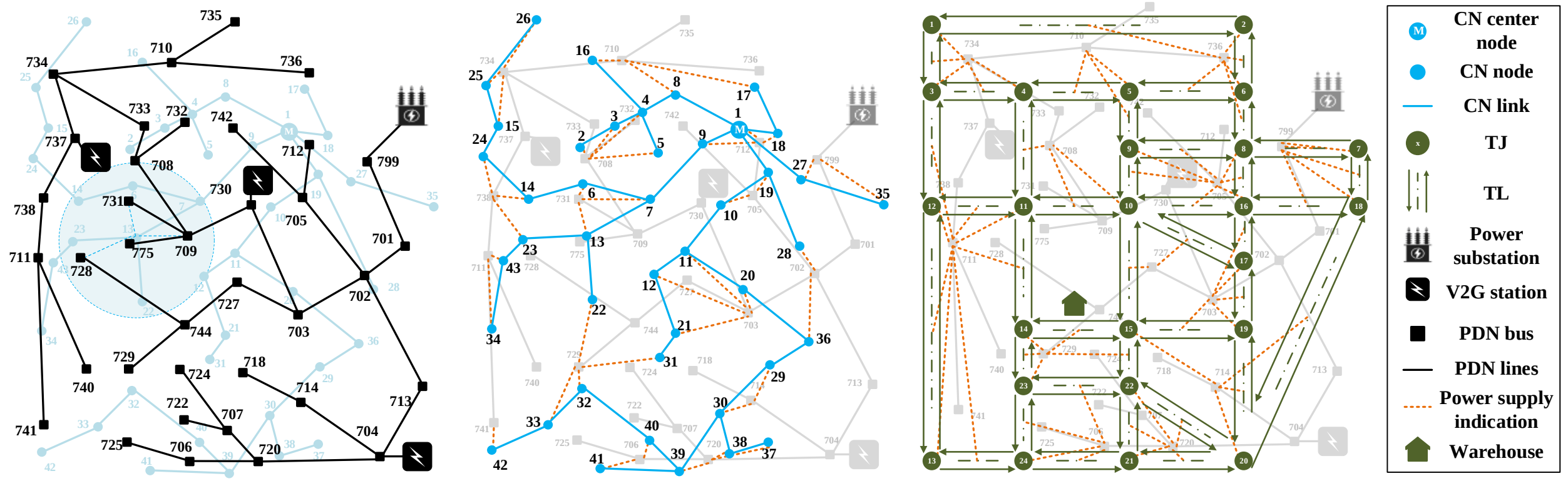

**Fig. 5.** Interactive support relations for the CN, UTN, and PDN in this simulation case

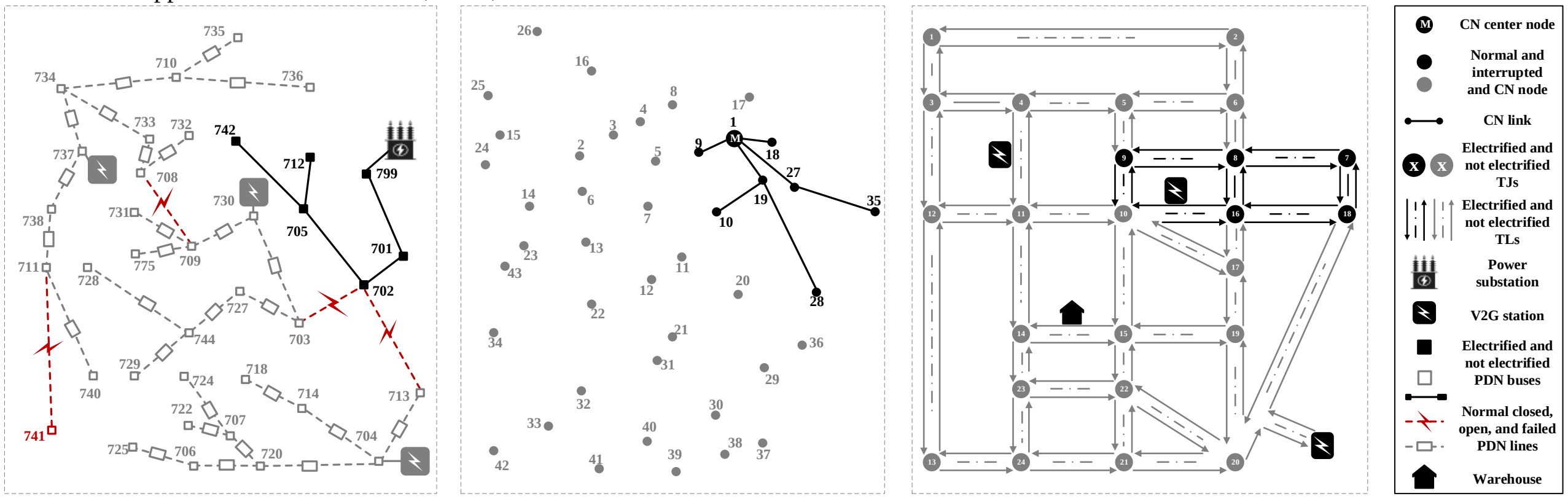

**Fig. 6.** The damage of disaster to this CN, UTN, and PDN integrated case

The complexity of integrated systems results from the diversity of subsystems. The proposed model and solution methods are directly verified on a modified IEEE 37-node test system, 24-node Sioux-Falls system, and 42-node wireless communication system integrated case with 3 V2GSs, as shown in Fig. 4. Fig. 5 shows the interdependencies of power supply and communication coverage among the CN, UTN, and PDN. Each base station is assumed to have a coverage area of 3 kilometers with 5 kW electricity demand. A UTN with approximately 2,600 stable operating vehicles is used in this case. Half of the on-road vehicles are EVs, and 30 percent of EVs are assumed to respond to V2G load regulation efforts based on EV users' willingness factor [30]. The speed limits of normal TJ areas and TLs are 30 km/h and 60 km/h, respectively. TLs are modeled using Sioux lane data that are five times shorter in length. The electricity demands of TL illumination and traffic signal lights at TJs are assumed to be 50 W/20 meters and 10 kW, respectively. The speeds of TLs and TJs within their immediate area are limited to 25 km/h and 3.3 km/h, respectively, when their facilities stop working.

The cases are calculated for illustrations and implemented on a normal PC with Intel Core i7 2.90 GHz and 32 GB memory. Yalmip on MATLAB 2021b with Gurobi 9.1.2 is utilized to solve the MILP problems, and the gap of MILP problems is set as 0.005. After testing, the objective function of $UDSSF$, which is frequently used in PDN line control capability judgment, can generally derive the results of UAV deployment in approximately 0.2~0.5 seconds. Therefore, we set a 10-second solution duration limitation for the solver. The UTN is simulated on the SUMO 1.18 software platform. TraCI4MATLAB [31] is used as the implementation of the interaction between MATLAB and SUMO.

The proposed algorithm (A3) and two other comparison algorithms (A1 and A2) are computationally validated on the same simulation platform and case. The first comparison algorithm (A1) schedules dispatchable MERSs to the V2GSs in the first stage to restore the load power supply in an upstream-to-downstream sequence. Then, MERSs are scheduled again to V2GSs to pick up loads as much as possible when the first stage of restoration work is completed. The second comparison algorithm (A2) schedules MESSs to V2GSs for UTN and CN load restoration in the first step. Then, MESRs are further dispatched in the second stage to maximize the power supply to PDN loads.

### *B. Disaster Scenario Generation*

In this case, four PDN line switches are at fault as a result of the disastrous event. The line switches that are in fault, open, or closed states are denoted by the red dashed, gray dashed, and black solid lines, respectively, in Fig. 6. Power outages cause a lot of communication and transportation facilities to stop working, resulting in a reduction in communication coverage and UTN speeds on a large scale. For this scenario, it is assumed that the warehouse has five MESSs with 500 kW/776 kWh capacity and 5 UAVs with 1-kilometer communication coverage, a fly speed of 180 km/h, and 50 kilometers of flight power parked on standby for emergency power and communication recovery. EVs are assumed to have the 50 kW/150 kWh capacity.

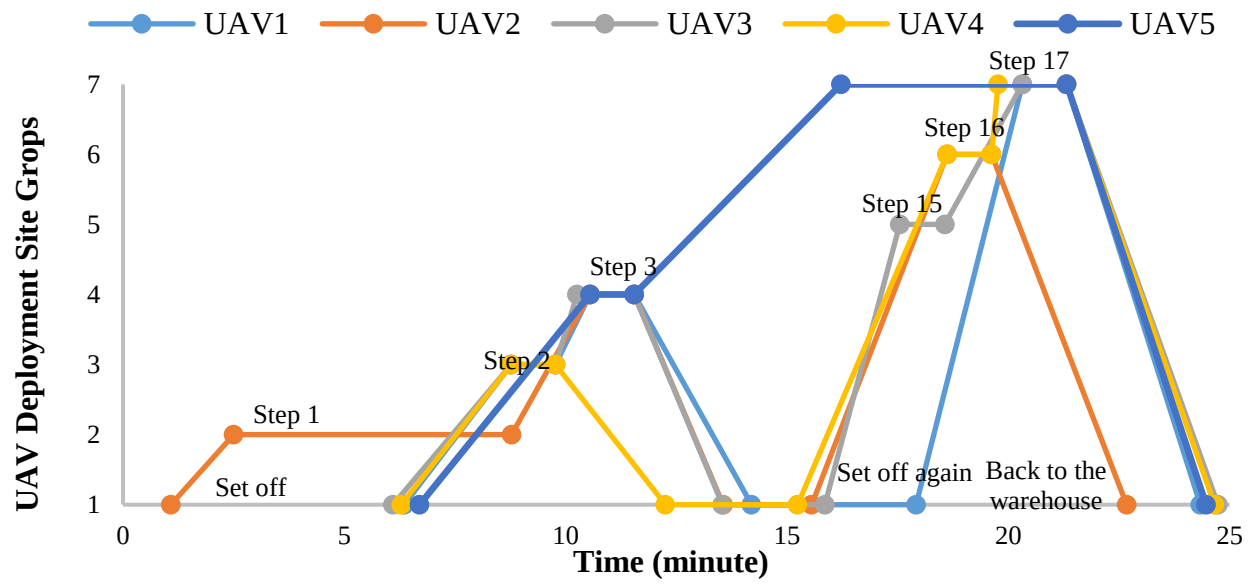

**Fig. 7.** UAV deployment site groups in the first-stage scheduling process of A3

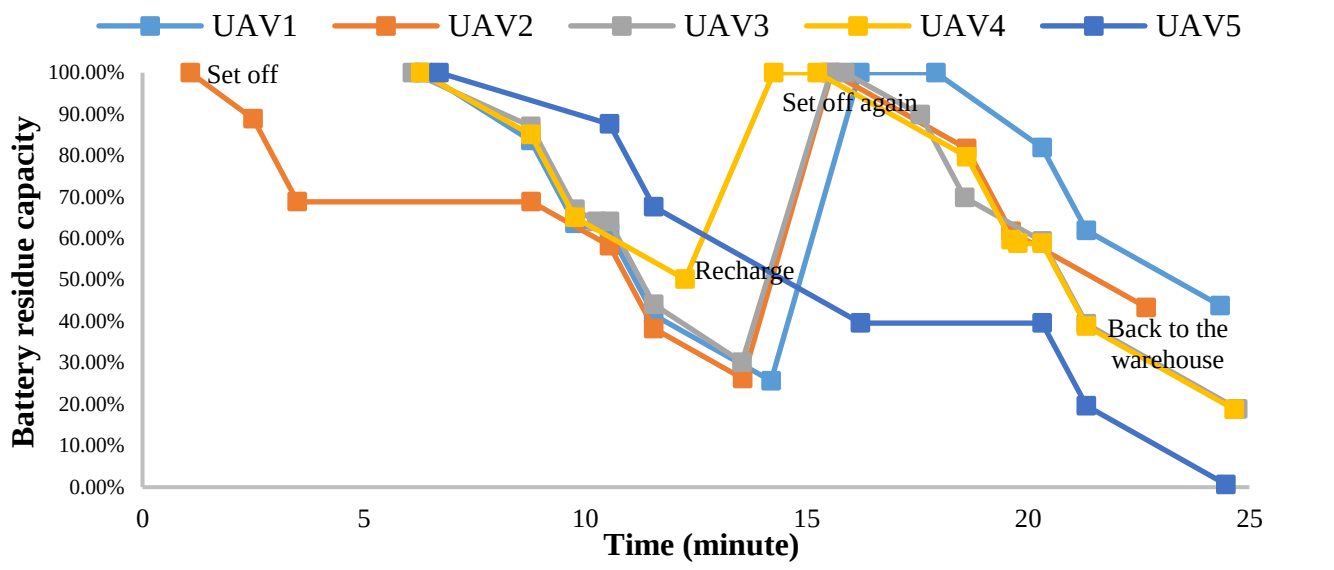

**Fig. 8.** UAV battery residue capacity in the first-stage scheduling process of A3

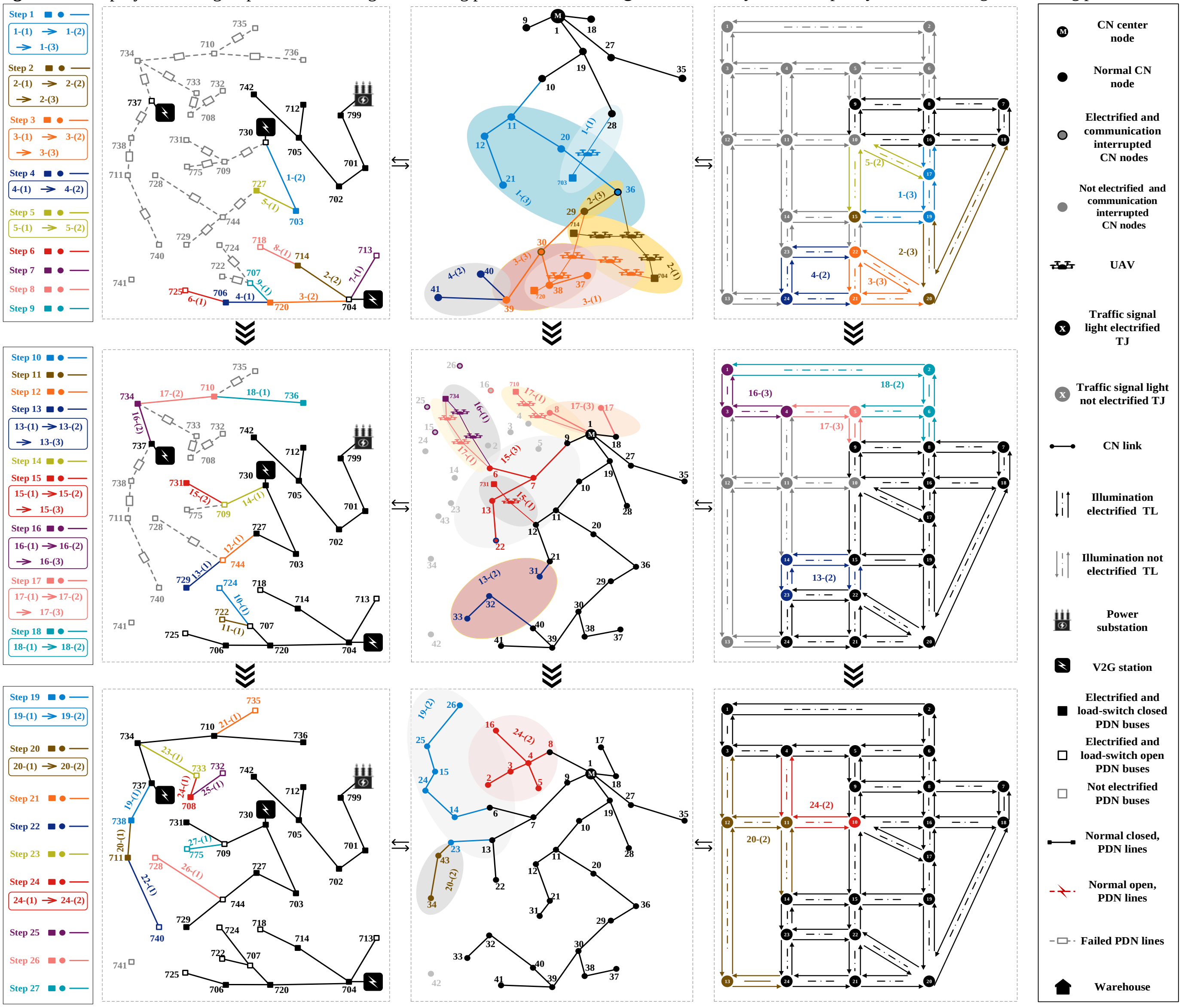

**Fig. 9.** Chained recovery process in the first-stage restoration of A3

## *C. Two-Stage Restoration Results*

Based on the above methodology, two-stage MESR-based restorations of these three algorithms are conducted. According to the calculation results of the first-stage algorithm in A3, the deployed flight position of UAVs and their residue capacity are shown in Figs. 7 and 8. Under the control of the first-stage algorithm, the. Fig. 9 shows the UTN, PDN, and CN interaction recovery process of A3. After 27 recovery steps, three microgrids energized by one V2GS are formed by opening/closing PDN line switches. The computation time for A3 is 278.21 s. Moreover, the final load recovery results of A1, A2, and A3 are shown in Fig. 10. According to Fig. 10, by comparing the results of A1 and A3 with those of A2, it can be seen that the schemes considering interdependent factors and prioritizing the restoration of communication and transportation facilities can improve communication coverage and thus improve the final load restoration capability. The A1 method does not take into account communication and transportation load prioritization. Although it can restore a small amount of load in a relatively short period, its load recovery capability is limited by the communication coverage and traffic conditions. Moreover, in A1, the DA remote control function is severely impaired by the relatively small communication coverage. Therefore, most subsequent recovery efforts are limited by the UAV dispatching process, further limiting the speed of load recovery. The capacity of the MESS is assumed to be sufficient to restore the power supply to communication and transportation facilities in this case. Therefore, when viewed as a whole, the load restoration curves of A2 and A3 are very close. This result illustrates the necessity of reserving an appropriate number of MESSs for urban PDNs. Most of the load recovery speeds in A3 are faster than those in A2, which illustrates the effectiveness of the proposed method in this paper. In addition, recovery of the communication and transportation loads after a disaster can also improve material transportation efficiency, resident travel convenience, and information interaction capability, thereby reducing economic losses from power outages.

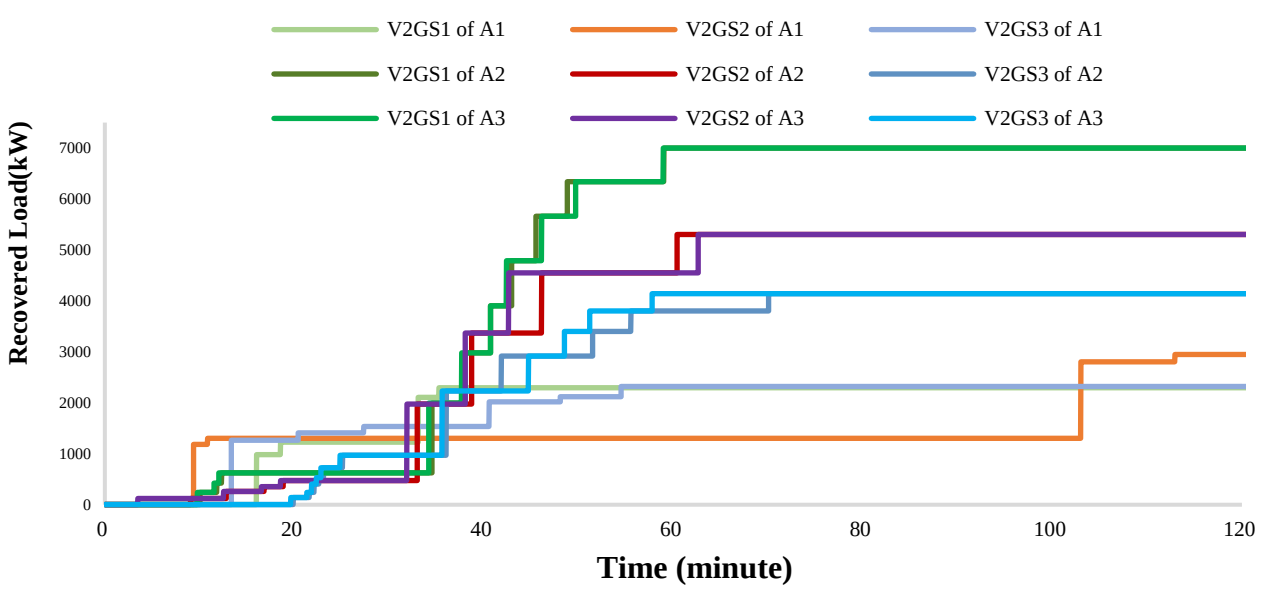

**Fig. 10.** Load recovery results of A1, A2, and A3

## VI. Conclusion

The interaction among PTIN infrastructures offers the potential for enhancing resilience in integrated systems. Prioritizing the power recovery power to facilities in the CN and UTN after a disaster can increase the capacity and speed of MESR-based PDN restoration. In this paper, we first explore and quantify the interacting factors of PTIN during MESR-based PDN restoration. Then, a PTN, UDN, and CN interaction model is formed to reflect the ‘chained recovery effect’ of the three networks. Given the importance of CN and UTN facilities in PDN restoration, a two-stage scheme is proposed, which restores CN and UTN loads in priority and then picks up other loads after the communication condition and traffic condition are improved. Better PDN restoration can be achieved with the proposed scheme. Case studies on the modified IEEE 37-node test system, 24-node Sioux-Falls system, and 42-node communication system integrated case highlight the merits of prioritizing the power recovery to communication and transportation facilities in comparison with the existing methods.

## References

[1] L. Zhang, C. Wang, J. Liang, M. Wu, B. Zhang, and W. Tang, “A coordinated restoration method of hybrid AC/DC distribution network for resilience enhancement,” *IEEE Trans. Smart Grid*, vol. 14, no. 1, pp. 112-125, Jan. 2022.

[2] C. Shao, M. Shahidehpour, X. Wang, X. Wang, and B. Wang, “Integrated planning of electricity and natural gas transportation systems for enhancing the power grid resilience,” *IEEE Trans. Power Syst.*, vol. 32, no. 6, pp. 4418-4429, Nov. 2017.

[3] A. H. Alobaidi, S. S. Fazlhashemi, M. Khodayar, J. Wang, and M. E. Khodayar, “Distribution Service Restoration With Renewable Energy Sources: A Review,” *IEEE Trans. Sustain. Energy*, vol. 14, no. 2, pp. 1151-1168, 2023.

[4] H. Gao, Y. Chen, S. Mei, S. Huang, and Y. Xu, “Resilience-oriented pre-hurricane resource allocation in distribution systems considering electric buses,” *Proc. IEEE*, vol. 105, no. 7, pp. 1214-1233, Jul. 2017.

[5] M. Nazemi, M. Moeini-Aghtaie, M. Fotuhi-Firuzabad, and P. Dehghanian, “Energy storage planning for enhanced resilience of power distribution networks against earthquakes,” *IEEE Trans. Sustain. Energy*, vol. 11, no. 2, pp. 795-806, Apr. 2019.

[6] T. Qian, C. Shao, X. Li, X. Wang, and M. Shahidehpour, ‘‘Enhanced coordinated operations of electric power and transportation networks via EV charging services,’’ *IEEE Trans. Smart Grid*, vol. 11, no. 4, pp. 3019-3030, Jul. 2020.

[7] L. Zhang, C. Wang, J. Liang, M. Wu, B. Zhang, and W. Tang, “A coordinated restoration method of hybrid AC/DC distribution network for resilience enhancement,” *IEEE Trans. Smart Grid*, vol. 14, no. 1, pp. 112-125, Jan. 2022.

[8] M. A. Igder, X. Liang, and M. Mitolo, “Service Restoration Through Microgrid Formation in Distribution Networks: A Review,” *IEEE Access*, vol.10, pp. 46618-46632, 2022.

[9] A. Sharma, D. Srinivasan, and A. Trivedi, “A decentralized multiagent system approach for service restoration using DG islanding,” *IEEE Trans. Smart Grid*, vol. 6, no. 6, pp. 2784-2793, Nov. 2015.

[10] S. Li, C. Gu, J. Li, H. Wang, and Q. Yang, “Boosting grid efficiency and resiliency by releasing V2G potentiality through a novel rolling prediction-decision framework and deep-LSTM algorithm,” *IEEE Syst. J*, vol. 15, no. 2, pp. 2562-2570, Jun. 2020.

[11] P. Jamborsalamati, M. J. Hossain, S. Taghizadeh, G. Konstantinou, M. Manbachi, and P. Dehghanian, “Enhancing power grid resilience through an IEC61850-based EV-assisted load restoration,” *IEEE Trans. Ind. Inform.*, vol. 16, no. 3, pp. 1799-1810, Mar. 2020.

[12] L. Zhang, C. Wang, J. Liang, M. Wu, B. Zhang, and W. Tang, “A coordinated restoration method of hybrid AC/DC distribution network for resilience enhancement,” IEEE Trans. Smart Grid, vol. 14, no. 1, pp. 112-125, Jan. 2022.

[13] B. Zhang, L. Zhang, W. Tang, Z. Wang and C. Wang, “A coordinated restoration method of electric buses and network reconfiguration in distribution systems under extreme events,” *CSEE J. Power Energy Syst.*, doi: 10.17775/CSEEJPES.2020.04320.

[14] J. Kim and Y. Dvorkin, ‘‘Enhancing distribution system resilience with mobile energy storage and microgrids,’’ *IEEE Trans. Smart Grid*, vol. 10, no. 5, pp. 4996-5006, Sep. 2019.

[15] S. Lei, C. Chen, H. Zhou and Y. Hou, "Routing and Scheduling of Mobile Power Sources for Distribution System Resilience Enhancement," *IEEE Trans. Smart Grid*,, vol. 10, no. 5, pp. 5650-5662, Sept. 2019.

[16] X. Wang, M. Shahidehpour, C. Jiang and Z. Li, "Resilience Enhancement Strategies for Power Distribution Network Coupled With Urban Transportation System," *IEEE Trans. Smart Grid*, vol. 10, no. 4, pp. 4068-4079, July 2019.

[17] A. Ferdowsi, A. Eldosouky and W. Saad, "Interdependence-Aware Game-Theoretic Framework for Secure Intelligent Transportation Systems," in *IEEE Internet Things J.*, vol. 8, no. 22, pp. 16395-16405, 15 Nov.15, 2021.

[18] S. V. Buldyrev, R. Parshani, G. Paul, H. E. Stanley, S. Havlin, “Catastrophic cascade of failures in interdependent networks,” *Nature*, vol. 464, no. 7291, pp. 1025-1028, Apr. 2010.

[19] K. Li, C. Shao, H. Zhang, and X. Wang, “Strategic pricing of electric vehicle charging service providers in coupled power-transportation networks,” *IEEE Trans. Smart Grid*, vol. 14, no. 3, pp. 2189–2201, May 2023.

[20] Z. Li, R. Al Hassan, M. Shahidehpour, S. Bahramirad and A. Khodaei, "A hierarchical framework for intelligent traffic management in smart cities", *IEEE Trans. Smart Grid*, vol. 10, no. 1, pp. 691-701, Jan. 2019.

[21] S. Lei, C. Chen, Y. Li, and Y. Hou, ‘‘Resilient disaster recovery logistics of distribution systems: Co-optimize service restoration with repair crew and mobile power source dispatch,’’ *IEEE Trans. Smart Grid*, vol. 10, no. 6, pp. 6187-6202, Nov. 2019.

[22] W. Wei and D. Wu, “Interdependence between transportation system and power distribution system: A comprehensive review on models and applications,” *J. Mod. Power Syst. Clean Energy*, vol. 7, no. 3, pp. 433-448, May 2019.

[23] H. Zhang, C. Chen, S. Lei and Z. Bie, "Resilient Distribution System Restoration With Communication Recovery by Drone Small Cells," *IEEE Trans. Smart Grid*, vol. 14, no. 2, pp. 1325-1328, March 2023.

[24] Y. Kawamoto, T. Mitsuhashi and N. Kato, "UAV-Aided Information Diffusion for Vehicle-to-Vehicle (V2V) in Disaster Scenarios," *IEEE Trans. Emerg. Top. Comput.*, vol. 10, no. 4, pp. 1909-1917, 1 Oct.-Dec. 2022,

[25] M. Shahidehpour, Z. Li, and M. Ganji, ‘‘Smart cities for a sustainable urbanization: Illuminating the need for establishing smart urban infrastructures,’’ *IEEE Electrific. Mag.*, vol. 6, no. 2, pp. 16–33, Jun. 2018.

[26] P. A. Lopez, E. Wiessner, M. Behrisch, L. Bieker-Walz, J. Erdmann, Y.-P. Flotterod, R. Hilbrich, L. Lucken, J. Rummel, and P. Wagner, “Microscopic traffic simulation using SUMO,” *Proc. 21st Int. Conf. Intell. Transp. Syst. (ITSC)*, Nov. 2018, pp. 2575–2582.

[27] T. Ding, Y. Lin, G. Li, and Z. Bie, “A New Model for Resilient Distribution Systems by Microgrids Formation,” *IEEE Trans. Power Syst.*, vol. 32, no. 5, pp. 4145-4147, Sep. 2017.

[28] A. Tzanakaki, M. P. Anastasopoulos, and D. Simeonidou, “Converged optical, wireless, and data center network infrastructures for 5G services,” *J. Opt. Commun. Netw.*, vol. 11, no. 2, pp. A111-A122, 2019.

[29] C. Huang and X. Wang, “A Bayesian approach to the design of backhauling topology for 5G IAB networks,” *IEEE Trans. Mobile Comput.*, vol. 22, no. 4, pp. 1867-1879, April 2023.

[30] X. Xu et al., "Evaluating Multitimescale Response Capability of EV Aggregator Considering Users’ Willingness," *IEEE Trans. Ind. Appl.*, vol. 57, no. 4, pp. 3366-3376, July-Aug. 2021.

[31] Acosta, Andrés F., et al. “TraCI4Matlab: Enabling the Integration of the SUMO Road Traffic Simulator and Matlab\Textregistered Through a Software Re-Engineering Process.” Modeling Mobility with Open Data, Springer International Publishing, 2015, pp. 155–70.

[32] Z. Ye et al., "Boost Distribution System Restoration With Emergency Communication Vehicles Considering Cyber-Physical Interdependence," *IEEE Trans. Smart Grid*, vol. 14, no. 2, pp. 1262-1275, March 2023

[33] K. P. Inala and K. Thirugnanam, "Role of Communication Networks on Vehicle-to-Grid (V2G) System in a Smart Grid Environment," 2022 4th International Conference on Energy, Power and Environment (ICEPE), Shillong, India, 2022, pp. 1-5